\documentclass[preprint]{revtex4}  
\usepackage{amsmath}
\usepackage{amssymb}
\usepackage{amsbsy}
\usepackage{graphicx}
\usepackage{wasysym}
\usepackage[dvips]{epsfig}
\usepackage{psfrag}
\usepackage[ansinew]{inputenc}
\usepackage[T1]{fontenc}
\usepackage{lmodern}
\usepackage{color}
\usepackage[dvipsnames]{xcolor}

\newcommand{\ket}[1]{\left|{#1}\right\rangle}

\newcommand{\op}[1]{\mathsf #1}
\newcommand{\set}[1]{\mathcal{#1}}

\renewcommand{\vec}[1]{\mbox{\boldmath $ #1 $}}

\def\gu{\;{\lower0.3ex\hbox{$\buildrel > \over {\scriptstyle \sim}$}}\;}
\def\lu{\;{\lower0.3ex\hbox{$\buildrel < \over {\scriptstyle \sim}$}}\;}

\def\XXint#1#2#3{{\setbox0=\hbox{$#1{#2#3}{\int}$}
     \vcenter{\hbox{$#2#3$}}\kern-.5\wd0}}

\begin{document}

\title{The planar Multipole Resonance Probe: \\a functional analytic approach}
\author{M. Friedrichs}
\author{J. Oberrath}
\affiliation{Institute of Product- and Process Innovation,\\  
	Department of Economic Sciences,\\ Leuphana University Lüneburg, D-21339 Lüneburg, Germany}
\date{\today}

\begin{abstract}

\textit{Active Plasma Resonance Spectroscopy} (APRS) is a well known diagnostic method, where a radio frequency probe is immersed into a plasma and excites plasma oscillations. The response of the plasma is recorded as frequency dependent spectrum, in which resonance peaks occur. By means of a mathematical model plasma parameters like the electron density or the electron temperature can be determined from the detected resonances. 

The majority of all APRS probes have in common, that they are immersed into the plasma and perturb the plasma due to the physical presence of the probe. To overcome this problem, the \textit{planar Multipole Resonance Probe (pMRP)} was invented, which can be integrated into the chamber wall of a plasma reactor. 

Within this paper, the first analytic model of the pMRP is presented, which is based on a cold plasma description of the electrons. The general admittance of the probe-plasma system is derived by means of functional analytic methods and a complete orthonormal set of basis functions. Explicit spectra for an approximated admittance including a convergence study are shown. The determined resonance frequencies are in good agreement with former simulation results.

\end{abstract}
\maketitle

\section{Introduction}

A plasma occupies the natural ability to resonate near the electron plasma frequency $\omega_{\rm pe}$. This ability is the essential requirement for active plasma resonance spectroscopy (APRS). Coupling a radio frequency (rf) signal in the GHz range into the plasma via an electrical probe the frequency dependent system response can be recorded to detect resonances. By means of a mathematical model to describe these resonance phenomena plasma parameters like electron density can be calculated.

Many different designs of APRS probes were invented, which are cited and classified in reference \cite{lapke2011}. One class of APRS probes excite electrostatic resonances \cite{takayama1960, messiaen1966, waletzko1967, vernet1975, blackwell2005, sugai1999, scharwitz2009, lapke2008}, which occur below $\omega_{\rm pe}$ and can be described by a model of the probe-plasma system in electrostatic approximation. Many approaches to understand these resonance phenomena have been reported \cite{fejer1964, harp1964, crawford1964, dote1965, kostelnicek1968, cohen1971, tarstrup1972, aso1973, bantin1974, booth2005, walker2006, lapke2007, xu2009, xu2010, li2010, liang2011}. They have in common, that their models are based on a fluid dynamical description and they focus only on a specific design of a certain probe.

However, the whole class of electrostatic probes can also be described generally. Applying functional analytic (Hilbert space) methods, a general and geometry independent solution of the system response can be derived \cite{lapke2013}. This response is identified as the electrical admittance of the probe-plasma system. Based on this solution one can proof, that the Multipole Resonance Probe (MRP) has the optimal design \cite{lapke2013, oberrath2014}.  

Apart from this fact, it is not suited for many industrial applications, because it is immersed into the plasma and disturbs it due to its physical presence. For this purpose a planar version of the MRP, the so called planar Multipole Resonance Probe (pMRP), was invented \cite{cschulz2014}. First measurements were compared to CST-simulations \cite{christianschulzpmrp} and the advantage of its planar design was shown.

However, to determine plasma parameters from the measured resonance peaks, a mathematical model is needed. In this manuscript the general description of APRS in electrostatic approximation will be applied to the geometry of the pMRP and an analytic solution for the admittance of the probe-plasma system will be presented. To determine specific spectra, the analytic solution has to be approximated, which requires a convergence study for different parameters. The finally converged spectra lead to a proportional relation between the resonance and plasma frequency. 

\pagebreak

\section{Model of the pMRP and its general admittance}\label{sec:ModelPMRP}

In a recent work a functional analytic description of APRS in electrostatic approximation for probes in arbitrary geometry was derived and a general solution of the admittance was presented \cite{oberrath2014}. In this section we present the model of the pMRP and apply the general description to its specific geometry. 

\begin{figure}[h]
\centering
\includegraphics[height=6cm]{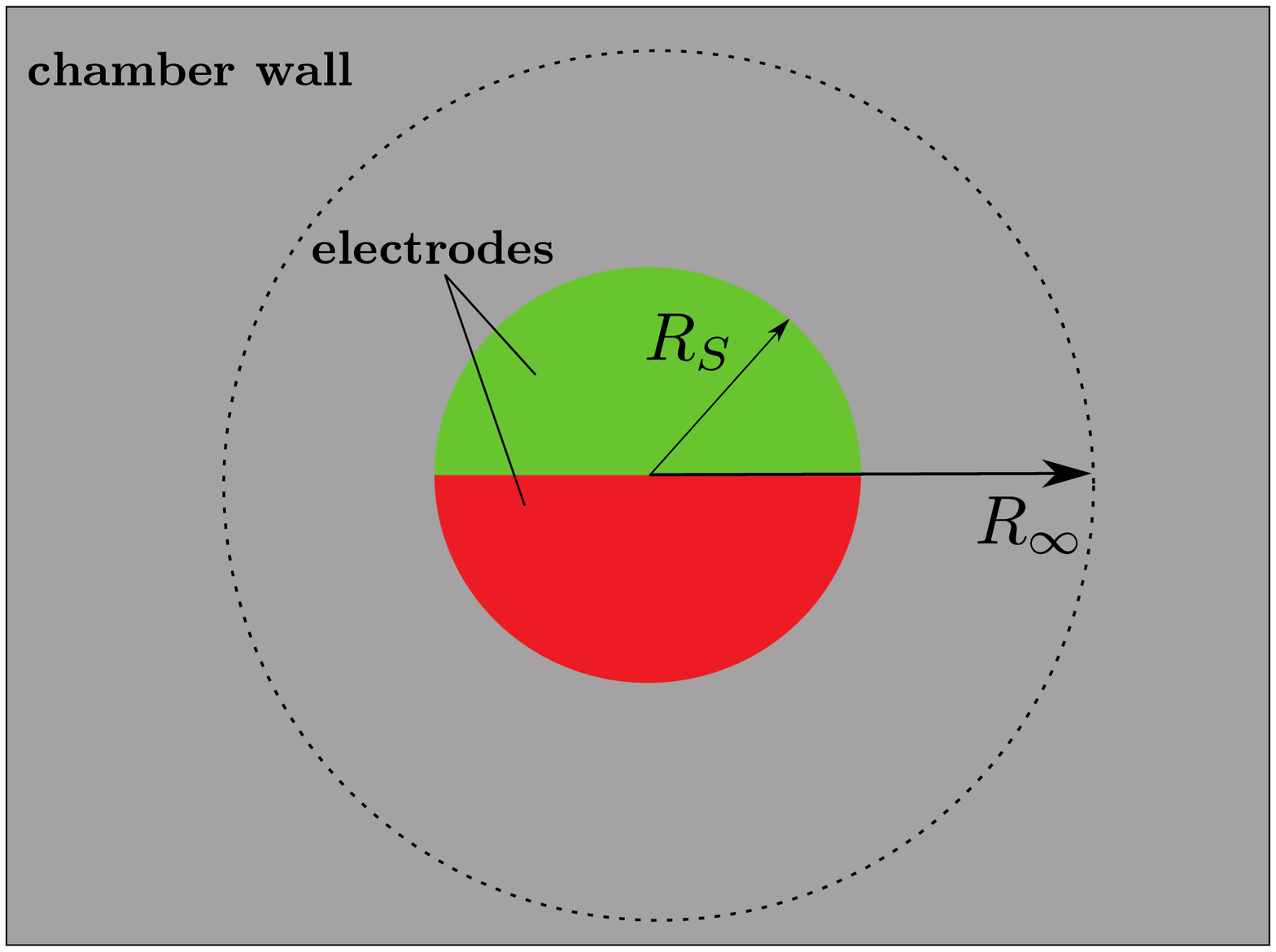}
\hspace{2mm}
\includegraphics[height=6.8cm]{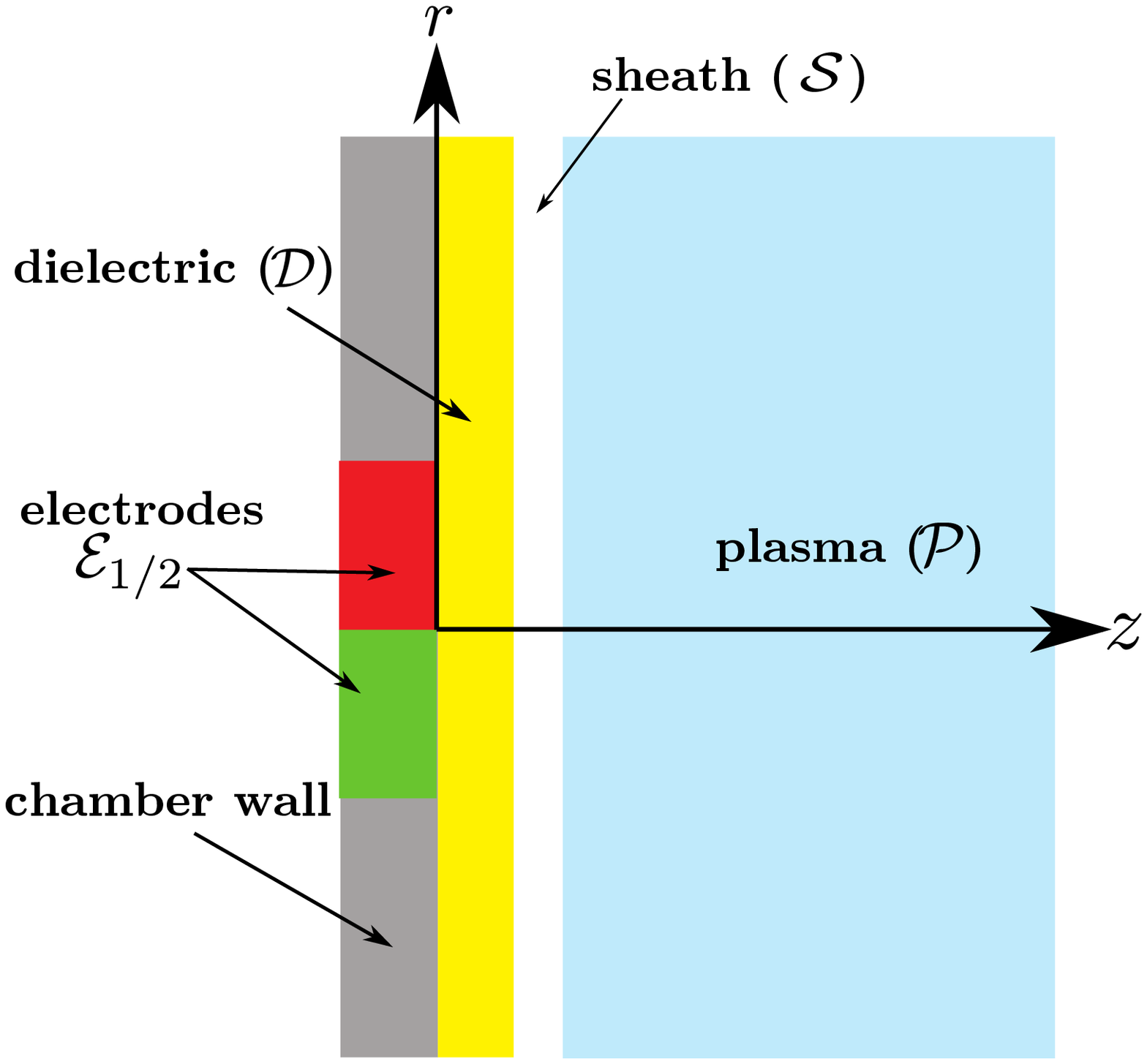}
\caption{The pMRP consists of two circular half disc electrodes $\set E_{1/2}$ with radius $R_{\textmd{S}}$ integrated into the chamber wall. A dielectric $\mathcal{D}$ covers the electrodes and the chamber wall. The sheath $\mathcal{S}$ in front of the dielectric separates it from the plasma $\mathcal{P}$.}
\label{fig:pmrpdraufsicht}
\end{figure}

As depict in fig.~\ref{fig:pmrpdraufsicht} the pMRP consists of two circular half disc electrodes $\mathcal E_{1/2}$ with the radius $R_S$, which are perfectly integrated into the chamber wall. The electrodes are insulated to each other and to the grounded chamber wall. A rf signal is applied to each of the electrodes, but with a 180 degree phase shift to each other. To allow for analytic solutions, a dielectric $\mathcal{D}$ with thickness $d$ covers the probe and the chamber wall (In reality only the electrodes are covered by the dielectric and the whole probe including the dielectric is integrated into the chamber wall.).    


$\mathcal{V}=\mathcal{P}\cup\mathcal{S}\cup\mathcal{D}$ is the domain of the dynamical interaction between the probe and the plasma $\mathcal{P}$, where $\mathcal{S}$ is the sheath with thickness $\delta$. Caused by the circular half disc electrodes a cylindrical coordinate system is chosen. The boundary $\partial\mathcal{V}$ of the dynamic interaction domain is then given by the electrode and wall surfaces at $z=0$ and at the surfaces $z\rightarrow\infty$ and $R=R_\infty$, where the dynamic interaction is assumed to vanish.   

\pagebreak

Within $\set S$ and $\set P$ the neutral gas is assumed as stationary background. Same holds for the ions, because the frequency $\omega$ of the applied signal is much larger than the ion-plasma-frequency $\omega_{\rm pi}$ ($\omega_{\rm pi}\ll \omega$). Thus, the dynamical behavior of the plasma can be described by the cold plasma model in electrostatic approximation for the electrons, given by the continuity equation and the generalized Ohm's law 
\begin{align}
\frac{\partial \sigma_{\textmd{e}}}{\partial t} &= -\left.\frac{}{}\vec{n} \cdot \vec{j_{\textmd{e}}}\right|_{z=d+\delta}\,,\\[2ex]
\frac{\partial \rho_{\textmd{e}}}{\partial t} &= -\nabla \cdot \vec{j_{\textmd{e}}}\,,\\[2ex]
\frac{\partial \vec{j_\textmd{e}}}{\partial t} &= -\varepsilon_0 \omega_{\textmd{pe}}^2\nabla\phi - \nu \vec{j_{\textmd{e}}}-\varepsilon_0 \omega_{\textmd{pe}}^2\sum_{k=1}^{2}U_k\nabla\psi_k\,,
\end{align}
The physical variables are charge density $\rho_{\textmd{e}}$ and current density $\vec{j_{\textmd{e}}}$ of the electrons. Due to the complete electron depletion within the sheath, a surface charge density $\sigma_{\textmd{e}}$ at the sheath edge $\mathcal{SK}$ has to be taken into account. $\phi$ is called inner potential and couples to the Poisson equation
\begin{equation}
-\nabla\cdot (\varepsilon_0\varepsilon_{\textmd{r}}\nabla\phi) 
= \left\{\begin{matrix}  0    & & \ \vec r  & \in & \set S \cup\set D \\[1ex] 
	\displaystyle \sigma_{\textmd{e}}    & & \ \vec{r} & \in & \set{SK} \\[1ex]
	\displaystyle\rho_{\textmd{e}}     & & \vec{r}   & \in & \set P
	\end{matrix}\right.\,
\end{equation}
with homogeneous boundary conditions. $\varepsilon_{\textmd{r}}$, the respective permittivity, is given as 1 within $\mathcal{P}$ and $\mathcal{S}$ and as $\varepsilon_{\textmd{D}}=const$ within $\mathcal{D}$.
The excitation of the plasma due to the rf signal is represented by the applied voltages $U_k$ and the characteristic functions $\psi_k$, which fulfill Laplace's equation
\begin{equation}
	\nabla^2\psi_k = 0
\end{equation}
with the boundary conditions
\begin{align}
	\left.\frac{}{}\psi_k(\vec{r})\right|_{\vec{r}\in\mathcal{E}_{k'}} &=\delta_{kk'} \,,\\[1ex]
	\left.\frac{}{}\psi_k(\vec{r})\right|_{z=0\,, r>R_{\textmd{S}}} &=0 \,,\\[1ex]
	\lim\limits_{z\rightarrow\infty} \psi_k(\vec{r}) &= 0\,.
\end{align}
$\delta_{kk'}$ is the Kronecker Delta, which equals 1 if $k=k'$ and 0 otherwise.  


As presented in \cite{oberrath2014}, the current at one electrode $\set E_1$ is defined as the scalar product of an excitation vector $\ket{e_1}=\left( 0 \,, 0 \,, -\varepsilon_0 \omega_{\textmd{pe}}^2\nabla\psi_1\right)^T$ and the dynamical state vector $ \left| z \right> = \left( \sigma_{\textmd{e}}\,,
\rho_{\textmd{e}}\,, \vec{j_{\textmd{e}}}\right)^{T} $ 
\begin{equation}
	i_1 = \left< e_1 \right|\left. z\right> 
	     = \sum_{k'=1}^{2} \left< e_1\right|\left.\left( i\omega-T_{\textmd{C}}-T_{\textmd{D}} \right)^{-1}\right.\left| e_{k'}\right>U_{k'} =\sum_{k'=1}^{2}Y_{1k'}U_{k'} \label{IPEP}\, .
\end{equation} 
Based on the general solution of the dynamical state vector, $Y_{1k'}$ can be identified as the coupling admittance between the electrodes. It is given by
\begin{equation}
	Y_{1k'} = \left< e_1\right|\left.\left( i\omega-T_{\textmd{C}}-T_{\textmd{D}} \right)^{-1}\right.\left| e_{k'}\right>\, ,
	\label{ExpCoupY}
\end{equation}
where $\op T_C$ and $\op T_D$ are the conservative and dissipative operator, respectively. They and the corresponding scalar product are defined in appendix \ref{sec:OpSca}.

\section{Expanded admittance of the pMRP}\label{sec:ExpAdm} 

To compute specific spectra of the pMRP, its coupling admittance has to be expanded in an appropriate complete orthonormal basis. Since the collision frequency $\nu$ in a low pressure plasma is much smaller than the frequency range of interest, a perturbation approach for operators can be applied and the set of eigenstate vectors $\{\ket{z_{nm}}\}$ of $\op T_C$ is a suitable choice \cite{oberrath2014} (To make this section more readable, all derivations are shifted to the appendix.). It can be derived by solving the eigenvalue equation
\begin{equation}
	T_{\textmd{C}}\ket{z_{nm}} = i\omega \ket{z_{nm}}\, .
	\label{eq:EigenvalueTC}
\end{equation}
To solve this eigenvalue problem in cylindrical coordinates we expand all scalar functions into cylindrical harmonics
\begin{align}
	\sigma_{\textmd{e}}(r,\varphi) &= \sum_{n=1}^{\infty}\sum_{m=0}^{\infty}\sigma_{nm}J_m(k_{nm}r)e^{im\varphi}\,,\\
	\rho_{\textmd{e}}(r,\varphi,z)  &= \sum_{n=1}^{\infty}\sum_{m=0}^{\infty} \rho_{nm}(z)J_m(k_{nm}r)e^{im\varphi}\,,\\
	\phi(r,\varphi,z) &= \sum_{n=1}^{\infty}\sum_{m=0}^{\infty}\phi_{nm}(z)J_m(k_{nm}r)e^{im\varphi}\,,
\end{align}
and the current density into vector cylindrical harmonics
\begin{equation}
	\vec{j_{\textmd{e}}}(r,\varphi,z) = \sum_{n=1}^{\infty}\sum_{m=0}^{\infty}\left( j_{nm}^{(R)}(z)\vec{R}_{nm} + j_{nm}^{(\Phi)}(z)\vec{\Phi}_{nm} + j_{nm}^{(Z)}(z)\vec{Z}_{nm} \right)\, .
\end{equation}
$J_m(k_{nm}r)$ represent Bessel's functions of the $m$-th order. $k_{nm}=j_{mn} R_{\infty}^{-1}$ is its $n$-th eigenvalue connected to the $n$-th root $j_{mn}$ of the $m$-th Bessel function. The vector cylindrical harmonics are orthonormal on circular surfaces with the radius $R_\infty$. Their definitions and some properties are shown in appendix \ref{sec:VCH}.

Similar to the calculations in \cite{oberrath2014}, the normalized eigenstate vector to the eigenvalue 
\begin{equation}
\omega_{nm} =\pm \frac{1}{\sqrt{2}}\sqrt{\left(1-\left(1-\frac{2}{\varepsilon_{\textmd{D}}\cosh(k_{nm}d)+1}\right)e^{-2k_{nm}\delta}\right)}\omega_{\textmd{pe}}
= \pm \eta_{nm}\omega_{\textmd{pe}}\label{EigVal}
\end{equation}
can be derived 
\begin{equation}
  \left|z_{nm}^{(\pm)}\right> 
= \left(\begin{matrix} \phi_{nm}(z)J_m(k_{nm}r)e^{im\varphi} \\[1ex] 
	\pm \varepsilon_0\omega_{\textmd{pe}}^2\frac{k_{nm}}{N_\textmd{Z}\omega_{nm}}\left[\vec{\Phi}_{nm} 
	- i \vec{Z}_{nm} \right]\phi_{nm}^{(\set{P})}(z)
	\end{matrix}\right)\,.\\
\end{equation}
It is important to note, that the inner potential is used in the eigenstate vector, but it couples unique to the surface charge density $\sigma_{\textmd{e}}$ and charge density $\rho_{\textmd{e}}$ via Poisson's equation. $N_{\textmd{Z}}$ represents the normalization coefficient of the vector cylindrical harmonics (see appendix \ref{sec:VCH}). To finalize the expansion of the coupling admittance, an explicit expression of the excitation vector is needed and is given as
\begin{equation}
		\left| e_k \right> = \left(\begin{matrix}
		 0\\  -\frac{\varepsilon_0\omega_{pe}^2}{N_{\textmd{Z}}}\left[ -\vec{\Phi}_{nm}ik_{nm}\psi_{nm}(z)+\vec{Z}_{nm}\frac{\mathrm{d}\psi_{nm}}{\mathrm{d}z} \right]\end{matrix} \right)\label{ExVec}\,,
\end{equation}
including the $z-$dependent part of the characteristic functions $\psi_{nm}$. 

Entering the completeness relation of the eigenstate vectors
\begin{equation}
	\sum_{n=1}^{\infty}\sum_{m=0}^{\infty}\left|z_{nm}^{(+)}\right>  \left<z_{nm}^{(+)}\right|+ \left|z_{nm}^{(-)}\right>  \left<z_{nm}^{(-)}\right|= 1
\end{equation}
twice and the excitation vector \eqref{ExVec} into \eqref{ExpCoupY}, the expanded coupling admittance of the pMRP reads as follows
\begin{equation}
Y_{1k'} = \sum_{n=1}^{\infty}\sum_{m=0}^{\infty} \left[ \frac{\varepsilon_0\omega_{\textmd{pe}}k_{nm}}{N_{\textmd{Z}}^2\eta_{nm}}B_{nm}^{(\set P)}e^{ -2k_{nm}\left( d+\delta \right) } \right]^2\frac{2i\omega \beta^{(1)}_{nm}\beta^{(k')}_{nm}}{\omega_{\textmd{pe}}^2\eta_{nm}^2+2i\omega\nu_{nm} -\omega^2}\label{ExpCoupYSimp}\, .	
\end{equation}
$B_{nm}^{(\mathcal{P})}$, $\beta^{(1)}_{nm}$, and $\beta^{(k')}_{nm}$ are integration constants, which are defined by the boundary and transition conditions of the eigenvalue problem (see appendix \ref{sec:eigVE} and \ref{sec:ExVe}). These constants include only geometric parameters of the probe. $\nu_{nm}=-\frac{1}{4}\nu$ are the matrix elements of the expanded dissipative operator $\op T_D$ (see appendix \ref{sec:AppTCTD}).

Finally, one can derive the admittance of the pMRP by entering the coupling admittance \eqref{ExpCoupYSimp} into \eqref{IPEP} and utilizing the 180 degree phase shift of the applied voltages $U_1=-U_2=U$:
\begin{equation}
Y =\displaystyle\sum_{n=1}^{\infty}\sum_{\tilde{m}=0}^{\infty}  
j_{2\tilde{m}+1,n}J_{2(\tilde{m}+1)}^2(j_{2\tilde{m}+1,n})e^{-2k_{n,2\tilde{m}+1}(d+\delta)}
\frac{4\pi\varepsilon_0\omega_{\textmd{pe}}^2 R_{\infty}i\omega\beta_{n,2\tilde{m}+1}^{(1)^2}}
	 {2\omega_{pe}^2\eta_{n,2\tilde{m}+1}^2-i\omega\nu-2\omega^2}\label{FinalY}\,.
\end{equation}
The admittance vanishes for even $m$, which yields a final sum over odd $m = 2\tilde{m}+1$.

\section{Converged spectra of the pMRP}\label{sec:ConSpec}

The admittance of the pMRP, derived within the last section, can be used to plot and analyze its spectrum. However, the spectrum is not given by an infinite number of discrete resonance modes, like the spectrum of probes with a spherical probe tip \cite{oberrath2014}, because the electrode geometry of the pMRP is not represented by Delta functions in the corresponding Fourier space. This means, that the admittance of the pMRP will have a spectrum with a broad resonance as a superposition of all addends in the double series \eqref{FinalY}. Thus, to determine explicit spectra, an approximated admittance with truncated sums is needed. \\

\begin{figure*}[h!]
	\centering
	\includegraphics[scale=0.55]{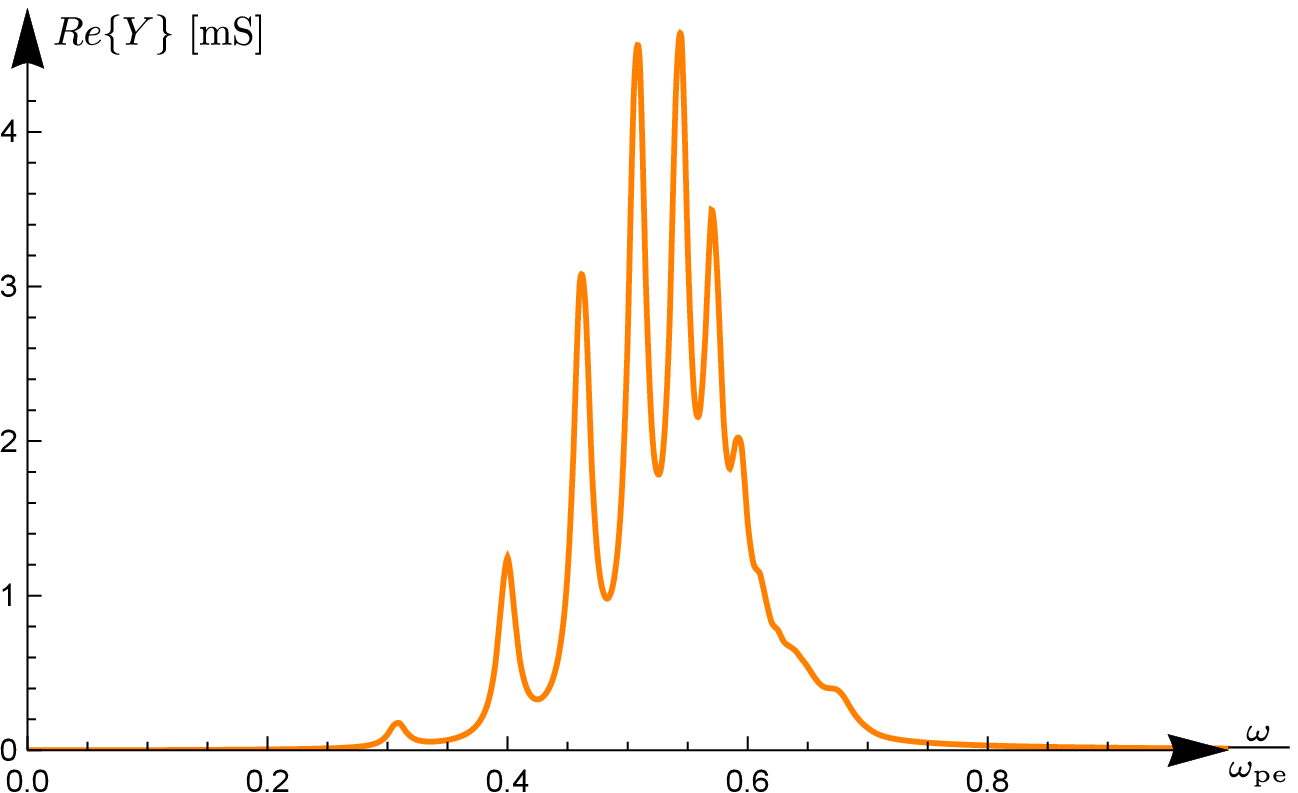}\hspace{2ex}
	\includegraphics[scale=0.55]{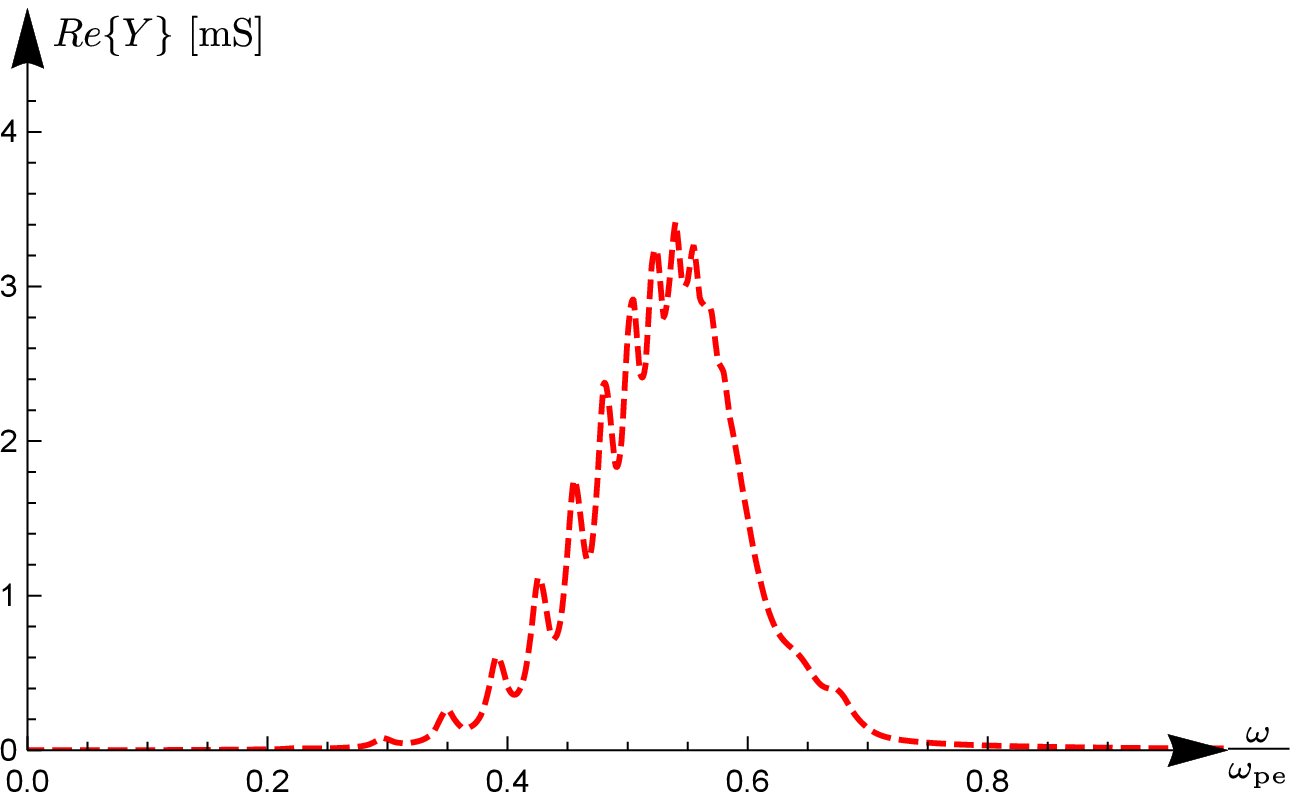}\\[1ex]
	\includegraphics[scale=0.55]{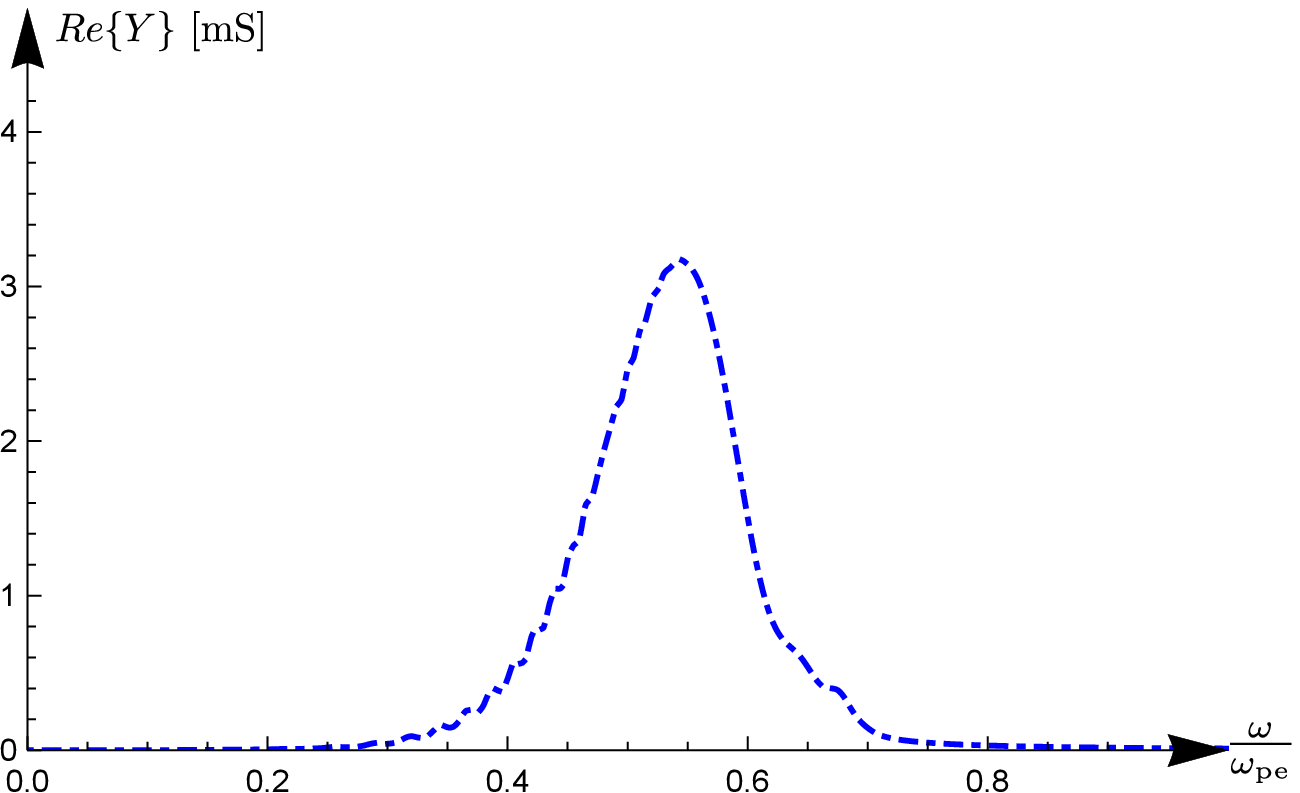}\hspace{2ex}
	\includegraphics[scale=0.55]{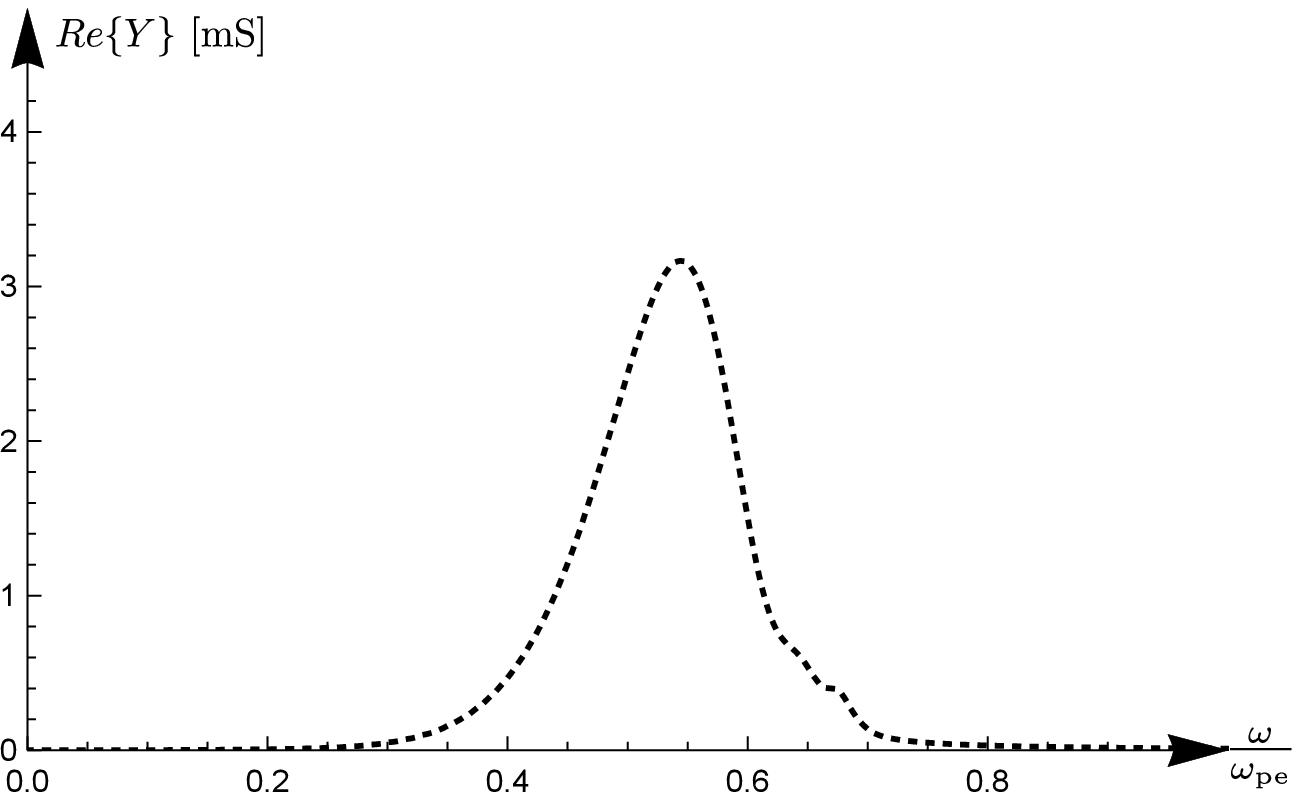}
	\caption{Spectra of the pMRP for constant $\tilde{M}_{\rm max}=10$ and $N_{\rm max}=200$ and varying radius \textcolor{orange}{$R_\infty=5\,R_{\textmd{S}}$ (bold)}, \textcolor{red}{$R_\infty=10\,R_{\textmd{S}}$ (dashed)}, \textcolor{blue}{$R_\infty=20\,R_{\textmd{S}}$ (dot-dashed)}, and $R_\infty=40\,R_{\textmd{S}}$ (dotted).}
	\label{fig:einflussrunendlich}
\end{figure*}
\pagebreak

Within this section we analyze the convergence behavior of the pMRP spectrum dependent on three parameters: the radius of the boundary surface $R_\infty$, the upper boundary of the inner sum $\tilde{M}_{\rm max}$ and the upper boundary of the outer sum $N_{\rm max}$. All other parameters are given by the geometry of the probe or the plasma itself and influence just the position of the resonance. In order to compare the determined spectra, we choose parameters of the pMRP taken from a recent published paper \cite{christianschulzpmrp}: $R_{\textmd{S}}= 3\,\textmd{mm}$, $d=0.4\,\textmd{mm}$, $\varepsilon_{\textmd{D}} =3.55$, $\delta =0.3 \,\textmd{mm}$, and $\nu=0.015\omega_{\rm pe}$.

Figure \ref{fig:einflussrunendlich} shows four different spectra to demonstrate the influence of $R_\infty$ for \textcolor{orange}{5 (bold)}, \textcolor{red}{10 (dashed)}, \textcolor{blue}{20 (dot-dashed)}, and 40 (dotted) times the probe radius $R_{\textmd{S}}$. $\tilde{M}_{\rm max}=10$ and $N_{\rm max}=200$ are set to large numbers, to focus only on the influence of $R_\infty$. If $R_\infty$ is too small, single peaks can be observed in the spectra. They correspond to certain eigenvalues, but have no physical meaning. The larger $R_\infty$ the smoother the spectrum gets and one broad resonance peak is formed. Above $R_\infty=40 R_{\textmd{S}}$ the spectrum is practically converged. Thus, $R_\infty=40 R_{\textmd{S}}$ represents the minimal radius for the boundary surface and will be used in further calculations within this manuscript. 

\begin{figure*}[h]
	\centering
	\includegraphics[scale=0.53]{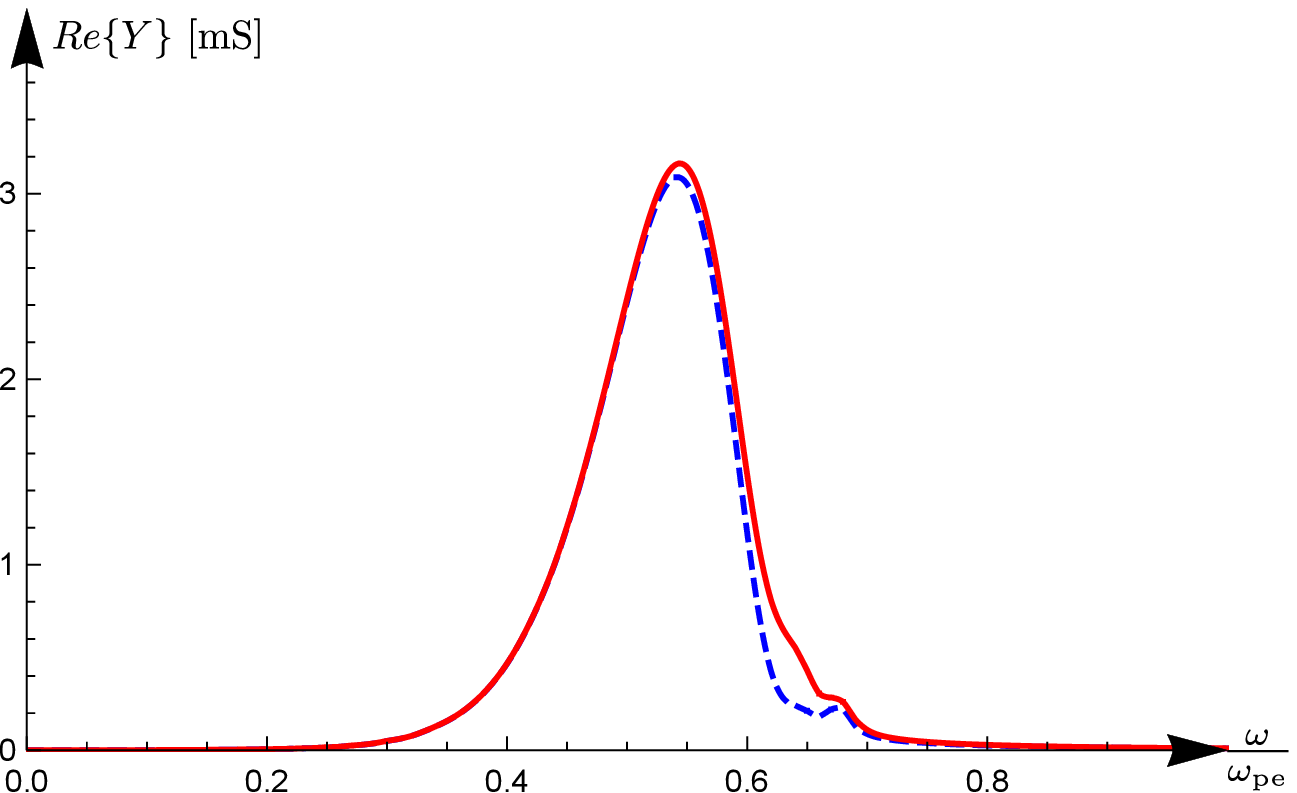}
	\hspace{5mm}
	\includegraphics[scale=0.53]{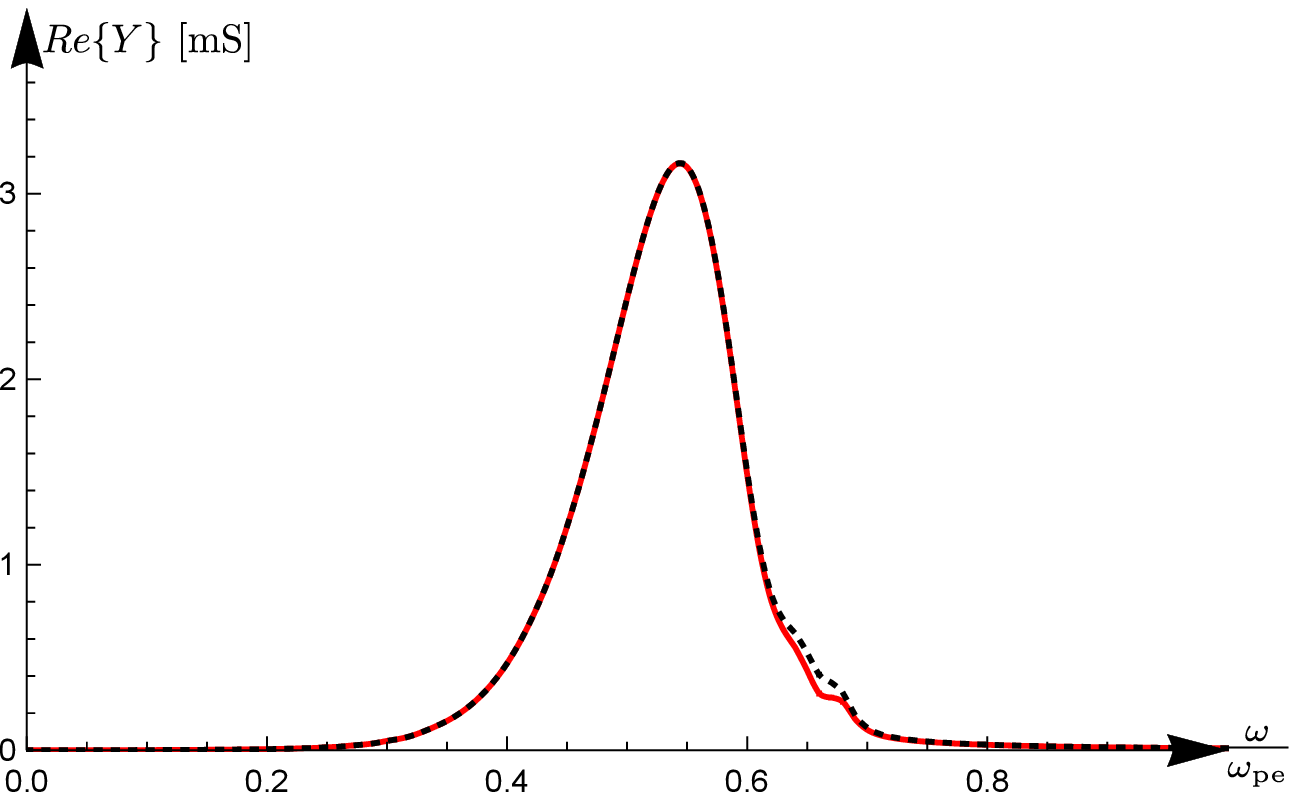}
	\vspace{-2ex}
	\caption{Spectra of the pMRP for constant $N_{\rm max}=200$ and $R_\infty=40\,R_{\textmd{S}}$ and varying \textcolor{blue}{$\tilde{M}_{\rm max}=0$ (dashed)}, \textcolor{red}{$\tilde{M}_{\rm max}=1$ (bold)}, and $\tilde{M}_{\rm max}=2$ (dotted).} 
		\label{fig:einflussmmax}
\end{figure*}

The influence of $\tilde{M}_{\rm max}$ is not tremendous. It is shown for $\tilde{M}_{\rm max}$ equal to 0, 1, and 2 with $N_{\rm max}=200$ in figure \ref{fig:einflussmmax}. A small difference in the height of the peaks and in the behavior above $\omega=0.6\,\omega_{\rm pe}$ can be observed in the spectra on the left hand side for \textcolor{blue}{$\tilde{M}_{\rm max}=0$ (dashed)} and \textcolor{red}{$\tilde{M}_{\rm max}=1$ (bold)}. Increasing $\tilde{M}_{\rm max}$ to 2 (see fig. \ref{fig:einflussmmax}, right, dotted) changes just the behavior above $\omega=0.6\,\omega_{\rm pe}$, which can be interpreted as contribution of higher modes, but the position and the height of the main peak remain the same. A further increase of $\tilde{M}_{\rm max}$ shows no difference within the spectra and can be neglected. Since the main resonance is not influenced by larger values of $\tilde{M}_{\rm max}$, we define it to 1 for further calculations.  

\begin{figure*}[h]
	\centering
	\includegraphics[scale=0.53]{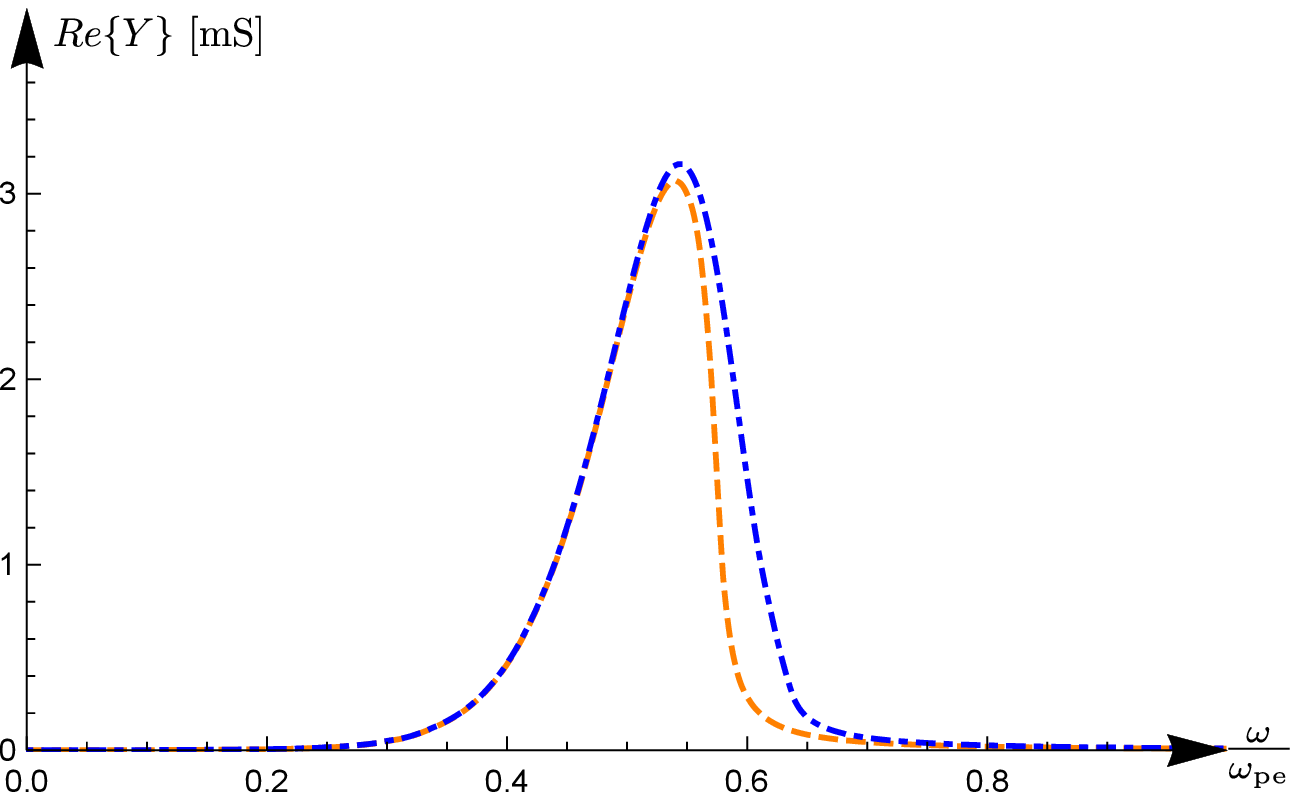}
	\hspace{5mm}
	\includegraphics[scale=0.53]{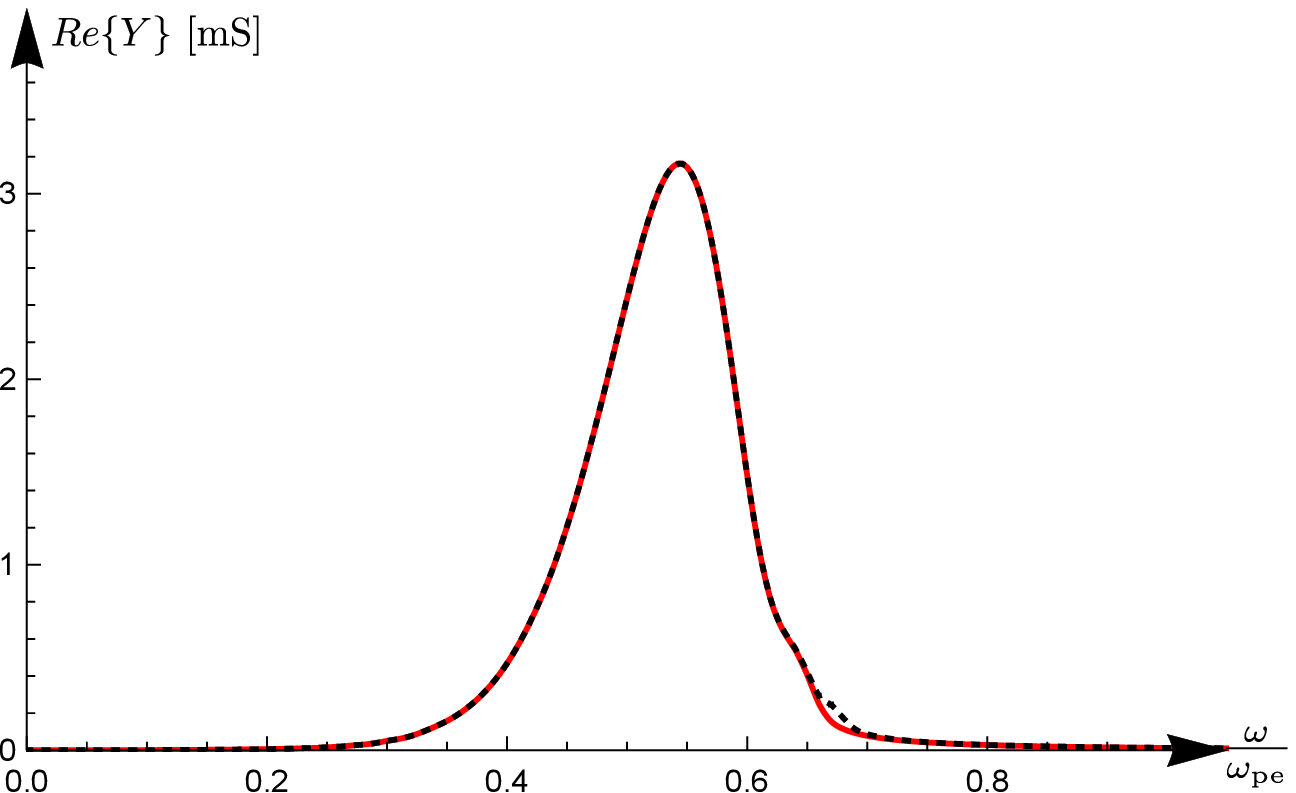}
	\vspace{-2ex}
	\caption{Spectra of the pMRP for constant $\tilde{M}_{\rm max}=1$ and $R_\infty=40\,R_{\rm s}$ and varying \textcolor{orange}{$N_{\rm max}=50$ (dashed)}, \textcolor{blue}{$N_{\rm max}=75$ (dot-dashed)}, \textcolor{red}{$N_{\rm max}=100$ (bold)} and $N_{\rm max}=125$ (dotted).}
	\label{fig:einflussnmax}
\end{figure*}

	
A strong influence is given by the outer sum truncated with $N_{\textmd{max}}$, which is shown in figure \ref{fig:einflussnmax}. On the left hand side the height and also the position of the peaks differ for $N_{\textmd{max}}$ equal to \textcolor{orange}{50 (dashed)} and \textcolor{blue}{75 (dot-dashed)}. The spectra on the right hand side for $N_{\textmd{max}}$ equal to \textcolor{red}{100 (bold)} and 125 (dotted) differ just slightly for values above $\omega=0.6\,\omega_{\rm pe}$, but height and position of the peak remain the same. A further increase of $N_{\textmd{max}}$ leads to identical spectra, which defines $N_{\textmd{max}}=125$ as smallest upper boundary of the outer sum for practically converged spectra. 

\begin{figure*}[h]
	\centering
	\includegraphics[scale=0.6]{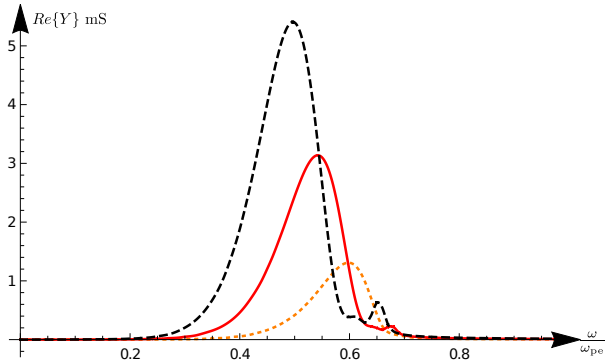}
	\vspace{-2ex}
	\caption{Converged spectra of the pMRP for different probe radii: \textcolor{orange}{$R_{\rm S}=2\,$mm (dotted)}, \textcolor{red}{$R_{\rm S}=3\,$mm (bold)}, and $R_{\rm S}=4\,$mm (dashed).}
	\label{fig:ConSpec}
\end{figure*}

Thus, practically converged spectra for the pMRP can be determined for $\tilde{M}_{\rm max}=1$, $N_{\rm max}=125$, and $R_\infty=40\,R_{\textmd{S}}$. In fig. \ref{fig:ConSpec} converged spectra for three different probe radii \textcolor{orange}{$R_{\rm S}=2\,$mm (dotted)}, \textcolor{red}{$R_{\rm S}=3\,$mm (bold)}, and $R_{\rm S}=4\,$mm (dashed) are depicted. The corresponding resonance frequencies are \textcolor{orange}{$\omega_{\rm r}=0.598\,\omega_{\rm pe}$}, \textcolor{red}{$\omega_{\rm r}=0.542\,\omega_{\rm pe}$}, and $\omega_{\rm r}=0.497\,\omega_{\rm pe}$ (dashed), respectively. They are in good agreement with the CST simulations presented in \cite{christianschulzpmrp} for the chosen probe and plasma parameters. 

\section{Conclusion}\label{sec:conclusion}

Within this work we presented the first analytic model of the pMRP and derived the general admittance of the probe-plasma system by means of functional analytic methods. To determine an explicit expression of the admittance a complete orthonormal set of basis functions was derived by the eigenvalue problem of the conservative operator $\op T_C$. They are based on cylindrical harmonic functions, due to the cylindrical geometry of the calculation domain. 

The explicit admittance is represented by an analytic expression, but it is given by an infinite expansion and has to be truncated to determine specific spectra of the pMRP. Due to that, a convergence analysis is presented to define the minimum values of the parameters $R_\infty=40R_S$, $N_{\textmd{max}}=125$, and $\tilde{M}_{\textmd{max}}=1$. They influence the convergence behavior. 

In the converged spectra a unique resonance peak is observable, similar to the spectra of the spherical impedance probe (sIP) or the MRP. However, its half width is much larger. This broader resonance peak is not caused by stronger damping or an additional damping mechanism like kinetic damping. It is due to the fact, that the eigenfunctions of the electrodes geometry are not represented by Delta functions in the corresponding Fourier space as in spherical geometry. 

Based on the converged spectra resonance frequencies for probes with different probe radii can be determined, which are in good agreement with former CST simulations \cite{christianschulzpmrp}. As proposed in this reference probes with different probe radii should be used to cover different frequency ranges. Within these frequency ranges the analytically determined resonance frequencies in this mansuscript are larger than that form the simulations. This is caused by a metallic adapter ring, which serves as planar impedance matching in the pMRP design. The simulations show, that a decrease of the inner radius of this ring shifts the resonances to larger frequencies, which leads to a better agreement with the analytic model.   


Based on the presented model the pMRP can be used for measurements of the electron density. It is an excellent candidate to monitor and/or control plasma processes. As a next step further detailed comparisons of our model in electrostatic approximation to full three dimensional electromagnetic simulations are planed to analyze the differences within the model and the simulation results. Furthermore, measurements with the pMRP compared to other diagnostic tools will follow.

\newpage
\appendix

\section{Operators and scalar product}\label{sec:OpSca}

The general definitions of the conservative and dissipative operator and the scalar product are given in \cite{lapke2013}. Here, we define these operators and the scalar product for the geometry of the pMRP. The operators are given by
\begin{eqnarray}
T_{\textmd{C}}\left| z \right> & = &
\left( -\left.\frac{}{}\vec{n} \cdot \vec{j_{\textmd{e}}}\right|_{z=d+\delta}\,, 
-\nabla \cdot \vec{j_{\textmd{e}}}\,, 
-\varepsilon_0 \omega_{\textmd{pe}}^2\nabla\phi \right)^T\ ,\\[1ex]
T_{\textmd{D}}\left| z \right> & = & \left( 0 \,, 0\,, -\nu\vec{j_{\textmd{e}}}\right)^T\ , 
\end{eqnarray}
and the scalar product between two different state vectors is defined as 
\begin{align}
\left< z' \right|\left. z \right> 
=& \int_{0}^{\infty}\int_{0}^{2\pi}\int_{0}^{R_\infty}
   \varepsilon_0\varepsilon_{\textmd{r}}\nabla\phi'^{*}\cdot \nabla\phi \,r\,
   \textmd{d}r\textmd{d}\varphi\textmd{d}z\\[1ex]
&+ \int_{d+\delta}^{\infty}\int_{0}^{2\pi}\int_{0}^{R_\infty}\frac{1}  
   {\varepsilon_0\omega_{\textmd{pe}}^2}\vec{j_{\textmd{e}}}'^{*}\cdot \vec{j_{\textmd{e}}}\,r\,  
   \textmd{d}r\textmd{d}\varphi\textmd{d}z\nonumber\, .
\end{align}

\section{Vector cylindrical harmonics}\label{sec:VCH}

The vector cylindrical harmonics used in this manuscript are defined as
\begin{align}
\vec{Z}_{nm} &= Z_{nm}\vec{e}_z = \frac{J_m(k_{nm}r)e^{im\varphi}}{\sqrt{\pi} R_\infty J_{m+1}(j_{mn})} \vec{e}_z\, = N_{\textmd{Z}}J_m(k_{nm}r)e^{im\varphi}\vec{e}_z\,,\\
	\vec{R}_{nm} &= \frac{1}{k_{nm}}\vec{L}Z_{nm}\,,\\
	\vec{\Phi}_{nm} &=\vec{e_z} \times \vec{R}_{nm}\,.
\end{align}
$\vec{L} = -i\, \vec{e_{z}}\times\vec{\nabla}$ is a rotation operator motivated by the angular momentum. These functions build an orthonormal basis on the circular surfaces with radius $R_\infty$ and fulfill the following orthogonal relations:
\begin{align}
	\int_{0}^{2\pi}\int_{0}^{R_\infty} \vec{Z_{nm}}^* \cdot \vec{Z_{n'm'}} \,r\textmd{d}r\textmd{d}\varphi &= \delta_{nn'}\delta_{mm'}\,,\\
	\int_{0}^{2\pi}\int_{0}^{R_\infty} \vec{R_{nm}}^* \cdot \vec{R_{n'm'}} \,r\textmd{d}r\textmd{d}\varphi &= \delta_{nn'}\delta_{mm'}\,,\\
	\int_{0}^{2\pi}\int_{0}^{R_\infty} \vec{\Phi_{nm}}^* \cdot \vec{\Phi_{n'm'}} \,r\textmd{d}r\textmd{d}\varphi &= \delta_{nn'}\delta_{mm'}\,,\\
	\int_{0}^{2\pi}\int_{0}^{R_\infty} \vec{Z_{nm}}^* \cdot \vec{R_{n'm'}} \,r\textmd{d}r\textmd{d}\varphi &= 0 \,,\\
	\int_{0}^{2\pi}\int_{0}^{R_\infty} \vec{Z_{nm}}^* \cdot \vec{\Phi_{n'm'}} \,r\textmd{d}r\textmd{d}\varphi &= 0 \,,\\
	\int_{0}^{2\pi}\int_{0}^{R_\infty} \vec{\Phi_{nm}}^* \cdot \vec{R_{n'm'}} \,r\textmd{d}r\textmd{d}\varphi &= 0 \,.
\end{align}

\section{Derivation of the eigenstate vector}\label{sec:eigVE}

The eigenvalue problem (\ref{eq:EigenvalueTC}) expanded into the scalar and vector cylindrical harmonics reads as follows
\begin{align}
i\omega\sigma_{nm} &= - j_{nm}^{(z)}\left.\frac{}{} \right|_{z=d+\delta}\,,\label{Dyn1}\\
i\omega\rho_{nm}&= \frac{\partial^2 \phi_{nm}}{\partial z^2}-ik_{nm}^2\phi_{nm}(z)\,,\\
i\omega j_{nm}^{(R)} &= 0\,,\\
i\omega j_{nm}^{(\Phi)} &= \frac{i\varepsilon_0 \omega_{\textmd{pe}}^2k_{nm}}{N_{\textmd{Z}}} \phi_{nm}\,,\\
i\omega j_{nm}^{(Z)} &= -\frac{\varepsilon_0\omega_{\textmd{pe}}^2}{N_{\textmd{Z}}}\frac{\partial\phi_{nm}}{\partial z}\label{Dyn5}\,.
\end{align}
Applying the expansion to Poisson's equation simplifies it to the $z-$component of the inner potential    
\begin{equation}
-\varepsilon_0\left[ \frac{\partial^2\phi_{nm}}{\partial z^2} - k_{nm}^2\phi_{nm} \right] = \left\{\begin{matrix} 0    &  &  z &\in & \set S \cup\set D \\
\displaystyle \sigma_e &  & z & \in & \set{SK} \\
\displaystyle\rho_e &  & z & \in & \set P
\end{matrix}\right.\label{ExpPoisson}\,
\end{equation}
with the boundary conditions
\begin{equation}
\phi_{nm}^{(\set{D})}(0) = 0 \qquad \textmd{and} \qquad \lim\limits_{z\rightarrow\infty} \phi_{nm}^{(\set{P})}(z) = 0\,,
\end{equation}
and the transition conditions
\begin{align}
\phi_{nm}^{(\set{D})}(d) - \phi_{nm}^{(\set{S})}(d) &= 0\,,\\
\phi_{nm}^{(\set{S})}(d+\delta)-\phi_{nm}^{(\set{P})}(d+\delta) &= 0\,,\\
\left[ \frac{\mathrm{d}\phi_{nm}^{(\set{S})}}{\mathrm{d}z} - \varepsilon_{\textmd{D}} \frac{\mathrm{d}\phi_{nm}^{(\set{D})}}{\mathrm{d}z} \right]_{z=d} &= 0\,,\\
\qquad \left[ \left( 1-\frac{\omega_{\textmd{pe}}^2}{\omega^2} \right) \frac{\mathrm{d}\phi_{nm}^{(\set{P})}}{\mathrm{d}z} -  \frac{\mathrm{d}\phi_{nm}^{(\set{S})}}{\mathrm{d}z} \right]_{z=d+\delta} &= 0\,.
\end{align}

Within the plasma domain $\set P$ the dynamic equations \eqref{Dyn1} to \eqref{Dyn5} can be simplified to one equation for the charge density $\rho_{nm}$, which can be entered into \eqref{ExpPoisson} to find
\begin{equation}
\left[ 1-\frac{\omega_{\textmd{pe}}^2}{\omega^2} \right]\left[ \frac{\mathrm{d}^2 \phi_{nm}^{(\set{P})}}{\mathrm{d}z^2} - k_{nm}^2\phi_{nm}^{(\set{P})} \right] = 0\label{ExpEigen}\,.
\end{equation}
Relevant frequencies have to fulfill $\omega \neq \omega_{\textmd{pe}}$, and thus the right bracket of equation \eqref{ExpEigen} has to be equal to zero, which represents the expanded Laplace equation. Also within the sheath $\set S$ and the dielectric $\set D$ Laplace's equation holds. Its general solution for the inner potential of all domains $\phi_{nm}^{(\set{D, S, P})}$ is given by
\begin{equation}
\phi_{nm}^{(\set{D},\set{S},\set{P})}(z) = A_{nm}^{(\set{D},\set{S},\set{P})}e^{k_{nm}z} + B_{nm}^{(\set{D},\set{S},\set{P})}e^{-k_{nm}z}\,.
\end{equation} 
Applying the transition and boundary conditions five of the six constants and the eigenvalues $\omega_{nm}$ of $T_{\textmd{C}}$ in cylindrical coordinates can be determined
\begin{align}
B^{(\set{D})}_{nm} &= -A_{nm}\,, \\
A^{(\set{S})}_{nm} &= \frac{1}{2}\left[ 1+e^{-2dk_{nm}} \left( \varepsilon_{\textmd{D}} -1 \right)+\varepsilon_{\textmd{D}} \right]A_{nm}\,, \\
B^{(\set{S})}_{nm} &= \frac{1}{2} \left[ -1-e^{2dk_{nm}}\left( \varepsilon_{\textmd{D}} -1 \right)-\varepsilon_{\textmd{D}} \right]A_{nm}\,, \\
A^{(\set{P})}_{nm} &= 0\,,\\
B^{(\set{P})}_{nm} &= -\frac{1}{2}\left[ (1+\varepsilon_{\textmd{D}})(1-e^{2k_{nm}\left( d+\delta \right)})+\left( \varepsilon_{\textmd{D}} -1 \right)\left( e^{2dk_{nm}}-e^{2\delta k_{nm}} \right) \right]A_{nm}\, .
\end{align}
$A_{nm}$, the last constant, can be derived by normalization of the eigenstate vector.

\section{Normalization}

The corresponding norm of an eigenstate vector is defined by the square root of the scalar product of two identical eigenstate vectors $\left| z_{nm}\right>$. In cylindrical geometry the scalar product of these eigenstate vectors reads as follows
\begin{align}
\left< z_{nm} \right|\left. z_{nm}\right> =& \frac{1}{N_{\textmd{Z}}^2}\int_{0}^{\infty}\varepsilon_0\varepsilon_{\textmd{r}}\left[ \left( \frac{\textmd{d}\phi_{nm}(z)}{\textmd{d}z} \right)^2 + k_{nm}^2\phi_{nm}^2(z) \right]\,\textmd{d}z\\
&+ \int_{d+\delta}^{\infty}\frac{1}{\varepsilon_0\omega_{\textmd{pe}}^2}\left[\left|j_{nm}^{(R)}(z) \right|^2 + \left|j_{nm}^{(\Phi)}(z) \right|^2 + \left|j_{nm}^{(Z)}(z) \right|^2 \right]\,\textmd{d}z \,\, .\nonumber
\end{align}
Fulfilling the condition $||z_{nm}||=\sqrt{\left< z_{nm} \right|\left. z_{nm}\right>}=1$, the last constant $A_{nm}$ can be determined
\begin{equation}
A_{nm}=\sqrt{ \frac{N_{\textmd{Z}}^2e^{-2k_{nm}d}\left[ \varepsilon_{\textmd{D}}-1+e^{2k_{nm}d}\left( \varepsilon_{\textmd{D}}+1 \right) \right]^{-1}}{ k_{nm}\varepsilon_0  \left[ (1+\varepsilon_{\textmd{D}})(1-e^{2k_{nm}\left( d+\delta \right)})+\left( \varepsilon_{\textmd{D}} -1 \right)\left( e^{2dk_{nm}}-e^{2\delta k_{nm}} \right) \right] } }\ .
\end{equation}

\pagebreak

\section{Derivation of the excitation vector}\label{sec:ExVe}

The excitation vector $\left| e_k \right> $ contains the characteristic functions $\psi_k$, which follow Laplace`s equation
\begin{equation}
	\nabla\cdot\left( \varepsilon_0\varepsilon_D\nabla\psi_k = 0 \right) \,\, \textmd{with} \,\, \lim\limits_{z\rightarrow\infty}\psi_k = 0  \,\, \textmd{and} \,\, \left.\psi_k\right|_{\set{E}_{k'}} = \delta_{kk'}\,.
\end{equation}
Similar to the derivation of the eigenstate vectors we expand the characteristic functions in cylindrical harmonics and determine the solution in $z-$direction:
\begin{equation}
	\psi_{nm} = \left\{\begin{matrix} \alpha^{(\set{D})}e^{k_{nm}z}  + \beta^{(\set{D})}e^{-k_{nm}z}  & & z & \in & \set D \\
	\displaystyle \beta^{(\textmd{vac})}e^{-k_{nm}z} &  & z & \in & \set S \cup\set P\,.
	\end{matrix}\right.
\end{equation}
The characteristic functions have to fulfill the following transition conditions:
\begin{equation}
	\psi^{(\set{D})}_{nm}(d) = \psi^{(\textmd{vac})}_{nm}(d) \qquad,\qquad\varepsilon_{\textmd{D}}\left.\frac{\partial}{\partial z}\psi^{(\set{D})}_{nm}\right|_d = \left.\frac{\partial}{\partial z}\psi^{(\textmd{vac})}_{nm}\right|_d\,.
\end{equation}
Based on these equations it is possible to determine the constants $\alpha^{(\set{D})}$ and $\beta^{(\set{D})}$ dependent on $\beta^{(\textmd{vac})} = \beta^{(k)}_{nm}$
\begin{align}
	\alpha^{(\set{D})} &= \frac{e^{-2k_{nm}d}\left(\varepsilon_{\textmd{D}}-1\right)}{2\varepsilon_{\textmd{D}}}\beta^{(k)}_{nm}\nonumber\ , \\
	\beta^{(\set{D})}&=\frac{\varepsilon_{\textmd{D}}+1}{2\varepsilon_{\textmd{D}}}\beta^{(k)}_{nm}\,.
\end{align}
The general excitation vector can then be determined to
\begin{equation}
	\left|e_k\right> = \sum_{n,m} \left( 0 \,,\, 0\,,\, \varepsilon_0\omega_{\textmd{pe}}^2\beta^{(k)}_{nm}k_{nm}e^{-k_{nm}z} \left[ \vec{\Phi}_{nm}i-\vec{Z}_{nm} \right] \right)\ .
\end{equation}
The remaining coefficient $\beta^{(k)}_{nm}$ can be calculated by the boundary condition $\psi_{nm}^{(k)}(0) = \delta_{kk'}$ at the electrodes $\set{E}_k$. Applying the orthogonality relation of the Bessel functions we find
\begin{align}
\beta^{(1)}_{nm} 
&= \frac{i\varepsilon_{\textmd{D}}\left( R_{\textmd{S}}N_{\textmd{Z}} \right)^2\left( \frac{k_{nm}R_{\textmd{S}}}{-2} \right)^m  \left( (-1)^m-1 \right) \Gamma(\frac{m}{2})\left._1F\right._2\left( \frac{m}{2}+1;\frac{m}{2}+2,m+1;-\frac{R_{\textmd{S}}^2 k_{nm}^2}{4} \right) }{  2\left[(\varepsilon_{\textmd{D}}-1)e^{-2k_{nm}d} +\varepsilon_{\textmd{D}}+1 \right]}\,\\[3ex]
\beta^{(2)}_{nm} \ ,
&=(-1)^m\beta^{(1)}_{nm} ,
\end{align}
for the pMRP. Here $\left._1F\right._2\left(a;\{b_1,b_2 \};z \right)$ is the generalized hypergeometric function and $\Gamma(z)$ the gamma function.

\pagebreak

\section{Matrix elements of the operators $T_C$ and $T_D$}
\label{sec:AppTCTD}

Due to the fact, that $\ket{z_{nm}}$ are the eigenstate vectors of the conservative operator $T_C$, its matrix elements can be calculated as
\begin{equation}
\left< z_{n'm'}\right|\left. T_C\right.\left|z_{nm}\right>=i\omega_{nm}\delta_{nn'}\delta_{mm'}\ .
\end{equation}
The dissipative operator applied to an eigenstate vector is given by
\begin{equation}
T_{\textmd{D}}\left|z_{nm}\right> = \left( 0 \,,\, 0\,,\, -\frac{\nu\varepsilon_0\omega_{\textmd{pe}}^2}{N_{\textmd{Z}}\omega_{nm}}\left[ \vec{\Psi}_{nm}k_{nm}\phi_{nm}^{(\mathcal{P})} + \vec{Z}_{nm}i\frac{\textmd{d}\phi_{nm}^{(\mathcal{P})}}{\textmd{d}z} \right] \right)
\end{equation}
and leads to
\begin{align}
\left<  z_{nm} \left|T_D\right| z_{n'm'} \right> 
&= -\frac{\nu (\varepsilon_0\omega_{\textmd{pe}}^2)^2}{ N_{\textmd{Z}}^2\omega_{n'm'}\omega_{nm}} {\displaystyle \int\limits_{d+\delta}^{\infty}} \frac{1}{\varepsilon_0\omega_{\textmd{pe}}^2} \left[ k_{n'm'}k_{nm}\phi_{n'm'}^{(\mathcal{P})}\phi_{nm}^{(\mathcal{P})} + \frac{\textmd{d}\phi_{n'm'}^{(\mathcal{P})}}{\textmd{d}z}\frac{\textmd{d}\phi_{nm}^{(\mathcal{P})}}{\textmd{d}z}  \right] \,\textmd{d}z\, \delta_{nn'}\delta_{mm'}\nonumber\\
&= -\frac{\nu\varepsilon_0\omega_{\rm pe}^2B_{nm}^{(\mathcal{P})^2}k_{nm}}{2N_{\textmd{Z}}^2\omega_{nm}^2}e^{-2k_{nm}(d+\delta)}=-\frac{\nu}{4}=\nu_{nm}\ .
\end{align}

\acknowledgments
The authors acknowledge support by the internal funding of the Leuphana University Lüneburg and the German Research Foundation via the project OB 469/1-1. Gratitude is expressed to J.~Gong, D.-B.~Grys, M.~Lapke, M.~Oberberg, D.~Pohle, C.~Schulz, J.~Runkel, R.~Storch, T.~Styrnoll, S.~Wilczek, T.~Mussenbrock, P.~Awakowicz, T.~Musch, and I.~Rolfes, who are or were part of the MRP-Team at Ruhr University Bochum. Explicit gratitude is expressed to R.P.~Brinkmann for fruitful discussions.

\end{document}